\documentclass[draft,jgrga]{agu2001}
\usepackage{lineno}

\authorrunninghead{Shaikh et al}

\titlerunninghead{Acoustic Gravity Wave Turbulence}

\authoraddr{
Dastgeer Shaikh,
 Institute of Geophysics and Planetary Physics (IGPP), University of California,
Riverside, CA 92521. USA. (dastgeer@ucr.edu)
}
\authoraddr{
Padma K. Shukla,  Institut f\"ur
Theoretische Physik IV and Centre for Plasma Science and Astrophysics,
Fakult\"at f\"ur Physik und Astronomie, Ruhr-Universit\"at Bochum,
D-44780 Bochum, Germany. (ps@tp4.rub.de) }

\authoraddr{
Lennart
Stenflo,  Department of Physics,
Umea University, SE-90187 Umea, Sweden
 (lennart.stenflo@physics.umu.se)}

\linenumbers*[1]

\newcommand{\be}{\begin{equation}}
\newcommand{\ee}{\end{equation}}

\newcommand{\Eqs}[2]{Equations  (\ref{#1}) and (\ref{#2})}

\newcommand{\Fig}[2]{Figs. (\ref{#1}) and (\ref{#2})}
\newcommand{\fig}[1]{Fig. (\ref{#1})}
 \newcommand{\eqa}{\begin{eqnarray}}
\newcommand{\eeq}{\end{eqnarray}}

\begin{document}

%
%
%
%
%

%
%

\title{Spectral Properties of Acoustic Gravity Wave Turbulence}
%

%
%


\author{Dastgeer Shaikh}
\affil{The Abdus Salam International Center for Theoretical Physics (ICTP),
Trieste, 34014 Trieste, Italy}

\author{Padma K. Shukla}
\affil{The Abdus Salam International Center for Theoretical Physics (ICTP),
Trieste, 34014 Trieste, Italy}

\author{Lennart Stenflo}
\affil{The Abdus Salam International Center for Theoretical Physics (ICTP),
Trieste, 34014 Trieste, Italy}

\begin{abstract}
The nonlinear turbulent interactions between acoustic gravity waves
are investigated using two dimensional nonlinear fluid
simulations. The acoustic gravity waves consist of velocity and
density perturbations and propagate across the density gradients in
the vertical direction in the Earth's atmosphere.  We find that the
coupled two component model exhibits generation of large scale
velocity potential flows along the vertical direction, while the
density perturbations relax towards an isotropic random distribution.
The characteristic turbulent spectrum associated with the system has a
Kolmogorov-like feature and tends to relax towards a $k^{-5/3}$
spectrum, where $k$ is a typical wavenumber. The cross field diffusion
associated with the velocity potential grows linearly and saturates in
the nonlinear phase.
\end{abstract}

%
%


%

\begin{article}

%
%
%

%
%


%
%

\section{Introduction}
Studies of acoustic gravity waves (AGWs) \citep{r1,r2,r3,r4,r5} in
the Earth's middle atmosphere, in the solar atmosphere, as well as
in planetary magnetospheres have been motivated by the need to
obtain accurate predictions of the dynamics of the atmosphere
under various meteorological conditions, and include the profiles
of the density, pressure, and the presence of shear flows in the
winds. The AGWs are  low-frequency disturbances associated with
the density and velocity perturbations of the atmospheric fluid in
the presence of the equilibrium pressure gradient that is
maintained by the gravity force. The frequency $\omega$ of the
AGWs is given by the dispersion relation \citep{r4,r5,r6}

\begin{equation}
\omega^2 =k_x^2 \omega_g^2/(k_x^2 + k_z^2+ 1/4H^2),
\end{equation}
where $k_x$ and $k_z$ are the components of the wave vector along the
$x$ and the $z$ axis in a Cartesian coordinated system, $H$ is the
density scale height, and the squared Brunt-V$\ddot{\rm a}$is$\ddot
{\rm a}$l$\ddot {\rm a}$ frequency is 
\be \omega_g^2 =
\left(\frac{d\rho_0^{-1}}{dz}+ \frac{1}{\gamma \rho_0
p_0}\frac{dp_0}{dz}\right)\left(\frac{dp_0}{dz}\right), 
\ee 
where $\rho_0 (z)$ and $p_0 (z)$ are the equilibrium density and pressure
that are inhomogeneous in the vertical direction (denoted by the
coordinate $z$), and $\gamma \approx 1.4$ is the adiabatic index.  At
equilibrium, we have $dp_0/dz =-\rho_0 g$, where $ {\bf g} =-g \hat
{\bf z}$ is the acceleration force on the atmospheric fluid due to
gravity, and $\hat {\bf z}$ is the unit vector in the vertical
direction. If $\omega_g^2 < 0$, it turns out from (1) that the AGWs
are unstable. When the amplitudes of the AGWs become large, the modes
start interacting among themselves. The mode couplings are governed by
\citep{r6}

\begin{equation}
\label{vel} \frac{D}{Dt}\left(\nabla^2 \psi
-\frac{1}{4H^2}\psi\right) =-\frac{\partial  \chi}{\partial x}
\end{equation}
and
\begin{equation}
\label{den}
\frac{D}{Dt}\chi =\omega_g^2 \frac{\partial \psi}{\partial x},
\end{equation}
where $\psi (x,z)$ is the stream function, $\chi (x,z)$ is the
normalized density perturbation, $D/Dt =(\partial/\partial t) + {\bf
  v}\cdot \nabla$, ${\bf v} =\hat {\bf z}(\partial \psi/\partial x)-
\hat {\bf x}(\partial \psi/\partial z)$ is the fluid velocity, and
$\nabla^2 = (\partial^2/\partial x^2) + (\partial^2/\partial
z^2)$. Time and space coordinates in Eqs. (3) \& (4) are normalized
respectively by their typical values $t_0$ and $\ell_0$. The other
variables are normalized as follows; $\psi/\psi_0=\psi',
~t_0\ell_0/\psi_0\chi=\chi', ~t_0^2\omega_g=\omega_g', ~H/\ell_0=H',
~\ell_0\nabla=\nabla', ~t_0D/Dt=D/Dt'$.  The primes have been removed
from Eqs. (3) \& (4).  Stationary nonlinear solutions of (3) and (4)
in the form of a double vortex and a vortex chain have been presented
by many authors \citep{r4,r5,r6,r7,r8}.

The coupled nonlinear equations conserve total energy $E$ in the
absence of sources and sinks. In the normalized units, it is
\[ E = \int [(\nabla \psi)^2 + (1/4H^2) \psi^2+\chi^2] dxdy. \]
It is noteworthy from the linear dispersion relation that the
group and phase speeds of the AGWs are negligibly small when $k_x
\gg k_z$. In such cases the fluctuations merely oscillate with the
Brunt-V$\ddot{\rm a}$is$\ddot{\rm a}$l$\ddot{\rm a}$ frequency. In
the other limit $k_x \ll k_z$, the AGWs are dispersive and  
have anisotropic propagation.  \Eqs{vel}{den} possess the
nonlinear terms $\hat{e}_y\times \nabla \nabla^2\psi \cdot
\nabla\chi $ and $\hat{e}_y\times \nabla\psi \cdot \nabla\chi $
that are analogous to the polarization and diamagnetic
nonlinearities in inhomogeneous drift wave turbulence. The former
cascades energy towards larger scales, while the latter leads to
the formation of smaller scales due to forward cascades. In the
next section, we shall explore the nonlinear interactions of
acoustic gravity waves.

The nonlinear interaction between finite amplitude AGWs in the Earth's
atmosphere \citep{gill} leads to energy transfer, resonantly and
non-resonantly between the high- and low-frequency parts of the
spectrum. Resonant interactions between the AGWs play also a key role
in a rotating atmosphere \citep{axelsson}. Many analytical theories
\citep{stenflo} describing nonlinear couplings between different
wavelength acoustic gravity modes predict the formation of vortices,
which can be important for weather predictions. Gardner
\citet{Gardner1994} explains vertical wave number spectrum by means of
a scale-independent diffusivity by assuming the damping effects of
molecular viscosity, turbulence, and off-resonance wave-wave
interactions. Similarly, \citep{beres2004} presented an implementation
of a physically based gravity wave source spectrum parameterization
over convection in Global Circulation Model.

With the objective of developing a self-consistent turbulent spectrum
of AGWs we, in this paper, present the turbulent properties of
nonlinearly interacting AGWs that are governed by Eqs. (3) and (4).
Specifically, we investigate by means of computer simulations the
statistical properties of the turbulent spectrum arising from dual
cascading. The fluid diffusion across the density and pressure
inhomogeneity directions in the presence of coherent structures is
also deduced. Furthermore, we consider the statistical turbulence
properties of the nonlinearly interacting AGWs.  Specifically, we
perform computer simulation studies of the acoustic gravity mode
cascading as well as the resulting structures and turbulent
spectra. The diffusion of the fluid across the density and pressure
gradients in the presence of large scale acoustic gravity mode
structures is also investigated.

\section{Nonlinear simulations}
To investigate the nonlinear mode coupling interaction and turbulence
aspects of acoustic gravity waves, we have developed a two dimensional
spectral code to numerically integrate equations (3) and (4). The 2D
simulations are not only computationally simpler and less expensive
compared with the full 3D simulations, but they offer significantly
higher resolution even on moderately sized small-cluster machines like
the IGPP (Institute of Geophysics and Planetary Physics, at University
of California Riverside) Beowulf. The spatial discretization in our
code uses a discrete Fourier representation of the turbulent
fluctuations.  The evolution variables use periodic boundary
conditions.  The initial isotropic turbulent spectrum is chosen close
to a $k^{-2}$ spectrum with random phases in all the directions. The
choice of such (or even a flatter than -2) spectrum treats the
turbulent fluctuations on an equal footing and avoids any influence on
the dynamical evolution that may be due to initial spectral
non-symmetry.  The equations are advanced in time using a Runge-Kutta
4 th order scheme. The code is made stable by a proper de-aliasing of
spurious Fourier modes and by choosing a relatively small time step in
the simulations. Our code is massively parallelized using Message
Passing Interface (MPI) libraries to facilitate higher resolution in a
2D computational box. The computational domain comprises $512^2$ modes
in two dimensions. Other simulation parameters in the normalized
units, as described above, are $H=10$, $\omega_g=0.02$, and $L=4
\times 4$ (two dimensional box size).

All the fluctuations in our simulations are initialized
isotropically with random phases and amplitudes in Fourier space.
The initial fluctuations do not possess any flows or mean fields.
The latter can be generated in our simulations by means of
nonlinear interactions. Fourier spectral methods are numerically
almost non dissipative compared to the existing finite difference
methods and are thus remarkably successful in describing turbulent
flows in a variety of plasma and hydrodynamic fluids.  They also
provide an accurate representation of the fluid fluctuations in
Fourier space. Because of the latter, nonlinear mode coupling
interactions preserve ideal rugged invariants of fluid flows,
unlike finite difference or finite volume methods. The
conservation of the ideal invariants such as energy, enstrophy,
magnetic potential, helicity etc. in the turbulence is an
extremely important feature because these quantities describe the
cascade of energy in the inertial regime, where turbulence is, in
principle, free from large-scale forcing as well as small scale
dissipation. We include small scale dissipation, however, to
distinctly extend the inertial range that helps push the spectral
cascades further down to the smallest scales. The latter leads to
an improved statistical average of spectral indices.  The
numerical validity in the simulations is checked by monitoring the
integrable form of the energy of the system.

The initially randomly propagating acoustic gravity waves begin to
interact linearly during the early phase of the simulations. As the
waves acquire larger amplitudes, they begin to interact
nonlinearly. During the nonlinear interaction phase, various eddies
mutually interact and transfer energy between the modes.  The
evolution of the modes for density fluctuations and velocity potential
fields are shown in \Fig{fig1}{fig2} in the $(x,z)$ plane.  It appears
that the density perturbations (i.e.  $\chi$ ) have a tendency to
generate smaller length-scale structures, while the velocity potential
cascades towards larger scales. This is consistent with the
corresponding nonlinear terms in drift wave turbulence.  The
co-existence of the small and larger scale structures in the
turbulence is an ubiquitous feature of various 2D turbulence
systems. For example, in 2D drift wave or CHM turbulence, the plasma
fluid admits two inviscid invariants, namely the energy and the mean
squared vorticity (i.e.  irrotational velocity fields). The two
invariants, under the action of an external forcing, cascade
simultaneously in turbulence, thereby leading to dual cascade
phenomena.

In these processes, the energy cascades towards longer
length-scales, while the fluid vorticity transfers spectral power
towards shorter length-scales. Usually a dual cascade is observed
in a driven turbulence simulation, in which certain modes are
excited externally through random turbulent forces in spectral
space. The randomly excited Fourier modes transfer the spectral
energy by conserving the constants of motion in $k$ space. On the
other hand, in freely decaying turbulence, the energy contained in
the large-scale eddies is transferred to  smaller scales leading
to a statistically stationary inertial regime associated with the
forward cascades of one of the invariants. Decaying turbulence
often leads to the formation of coherent structures as the
turbulence relaxes, thus making the nonlinear interactions rather
inefficient when they are saturated. It seems that the acoustic
gravity waves exhibit turbulence in a manner similar to two
dimensional hydrodynamic turbulence in which the density and
velocity potential fluctuations are analogous to the hydrodynamic
fluid energy and vorticity. The evolution of the turbulent energy
associated with the modes (density and potential field) are shown
in \fig{fig3} where the linear and nonlinear phases of the
evolution are clearly marked by the curve. After the nonlinear
interactions are saturated, the energy in the turbulence does not
grow and remains nearly unchanged throughout the simulations.
Correspondingly, the energy transfer rate shows a significant
growth during the linear and initial nonlinear phases. However
when the nonlinear interactions saturate, the nonlinear transfer
of energy in the spectral space amongst various turbulent modes
becomes inefficient and the energy transfer per unit time tends to
become negligibly small as shown in \fig{fig4}. The power spectrum
associated with the acoustic gravity wave turbulence exhibits a
spectral slope close to $-5/3$ as shown in \fig{fig5}. This is
indicative of the turbulent eddies being the dominant processes in
the spectral transfer and is consistent with the Kolmogorov-like
phenomenology.

We finally estimate the turbulent transport coefficient associated
with a self-consistent evolution of the small and large scale
turbulent fluctuations.  An effective electron diffusion
coefficient caused by the momentum transfer can be calculated from
$D_{eff} = \int_0^\infty \langle {\bf v}({\bf r},t) \cdot {\bf
v}({\bf r},t+ t^\prime) \rangle dt^\prime$, where ${\bf v}$ is the
fluid velocity, the angular bracket represents spatial averages
and the ensemble averages are normalized to unit mass. Since the
2D structures are confined to a $x-z$ plane, the effective
diffusion coefficient, $D_{eff}$, essentially relates the
diffusion processes associated with the translational motion of
the fluid elements  of our nonlinearly coupled system. We compute
$D_{eff}$ in our simulations from both the velocity field
fluctuations to measure the turbulent transport that is associated
with the large scale flows.  It is observed that the effective
cross-field transport is lower initially, when the field
perturbations are Gaussian. On the other hand, the diffusion
increases rapidly with the eventual formation of longer
length-scale structures. This is shown in \fig{fig6}.  The
transport due to the electrostatic potential dominates
substantially as depicted in the figure.  Furthermore, in the
steady-state, the nonlinearly coupled modes form stationary
structures and $D_{eff}$ saturates eventually. Thus, remarkably,
an enhanced cross-field transport level results primarily due to
the emergence of large-scale coherent structures in a turbulence
dominated by the acoustic gravity waves.

\section{Concluding Remarks}
In summary, we have investigated the turbulent properties of 2D
nonlinearly interacting AGWs by means of computer simulations. We
found that the nonlinear equations governing the dynamics of the
fluid vorticity and density fluctuations in a nonuniform fluid
admit dual cascade, leading to a Kolmogorov-like energy spectrum.
In such a cascade process, we observe the formation of the short
and large scale structures that are responsible for the
cross-field (across the density gradient direction) diffusion of
particles due to random walk processes. The present results are
essential for understanding the turbulent properties of the
atmospheres of the Earth and other planets.

%
%
%
%
%
%
%
%


\end{article}

%
%
%

\newpage

\begin{figure}
\caption{\label{fig1} Small scale fluctuations in the wave density
fluctuations as a result of  steady turbulence simulations of
two-dimensional acoustic gravity waves. Forward cascades are
responsible for the generation of small-scale fluctuations. The
density perturbations are distributed isotropically in the two-
dimensional domain.}
\end{figure}

\begin{figure}
\caption{\label{fig2}Evolution of potential fluctuations.
Formation of large scale structures along the direction of the
density inhomogeneity.}
\end{figure}

\begin{figure}
\caption{\label{fig3} Time evolution of the turbulent energy. The
initial linear phase is followed by a nonlinear phase. The
saturation follows at later times.}
\end{figure}

\begin{figure}
\caption{\label{fig4} Energy cascade rates. A rapid transfer of
energy takes place during the early phase of the evolution
followed by  nonlinear saturation. In the latter, the energy
cascade rates are saturated leading thereby to no net transfer
amongst the turbulent eddies.}
\end{figure}

\begin{figure}
\caption{\label{fig5} The AGW turbulence exhibits a
Kolmogorov-like $k^{-5/3}$ power spectrum in the inertial range.}
\end{figure}

\begin{figure}
\caption{\label{fig6} Time evolution of the effective diffusion
coefficient.}
\end{figure}


%

%
%

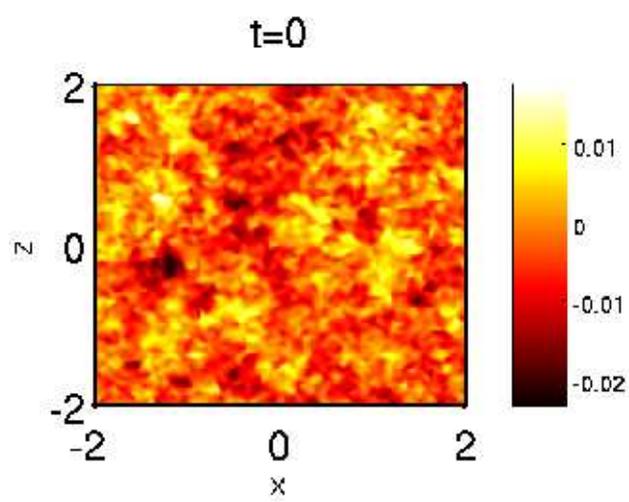
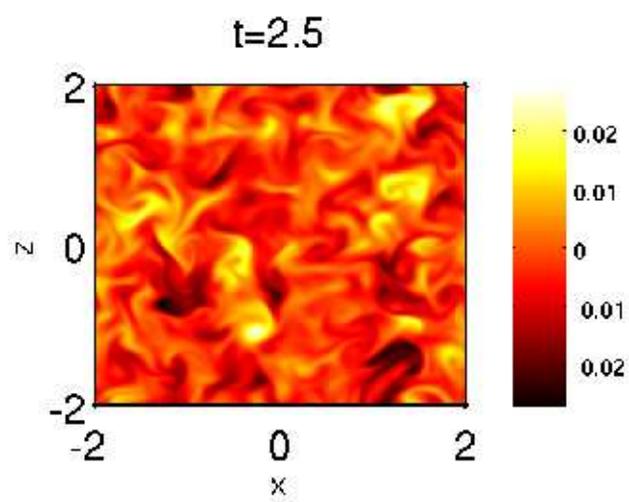
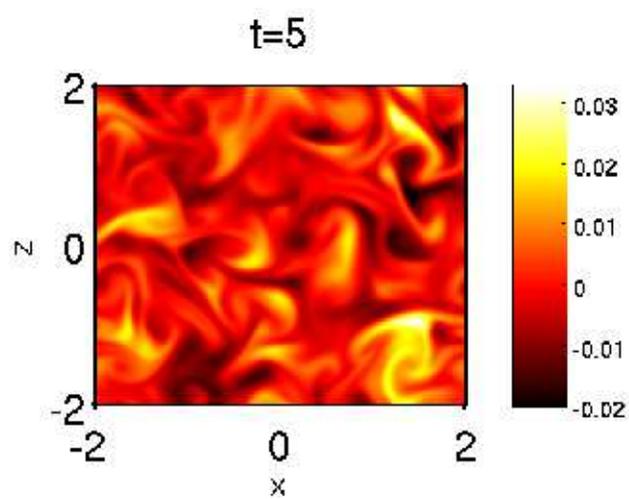
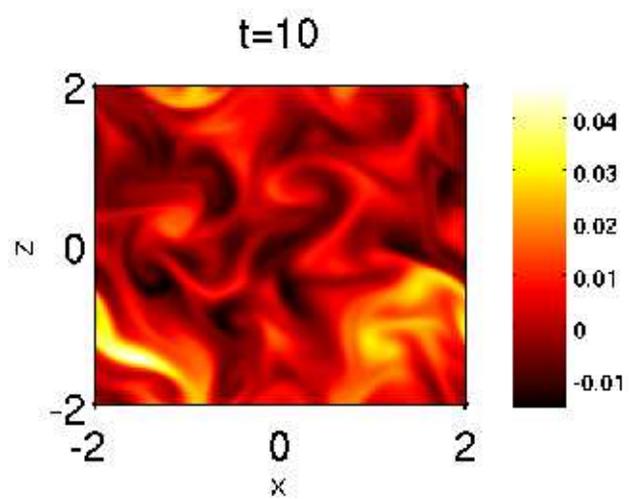
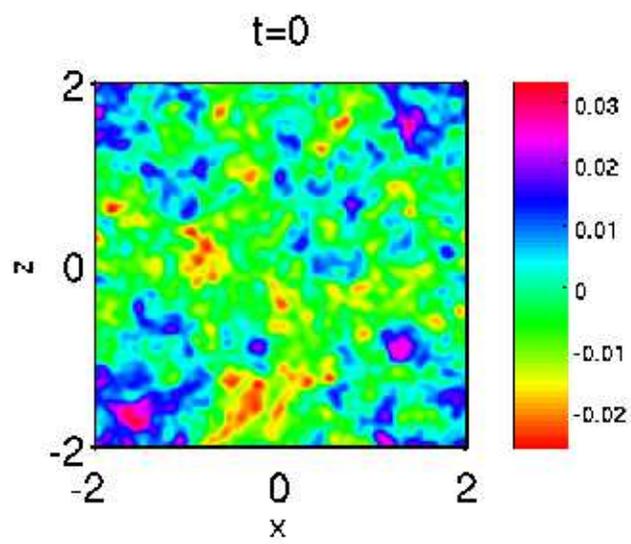
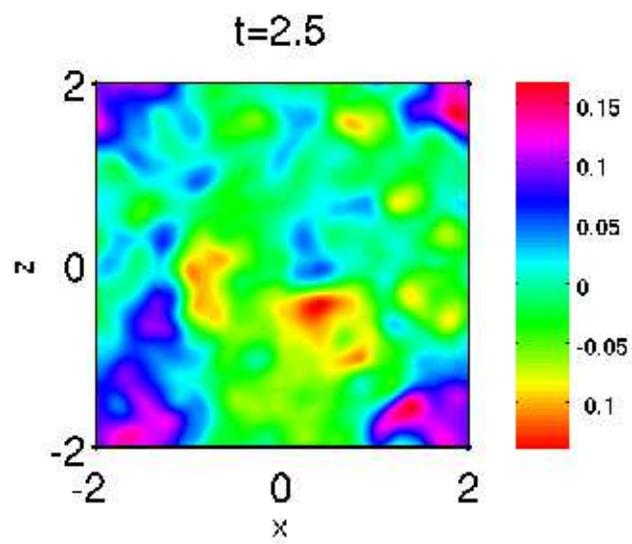
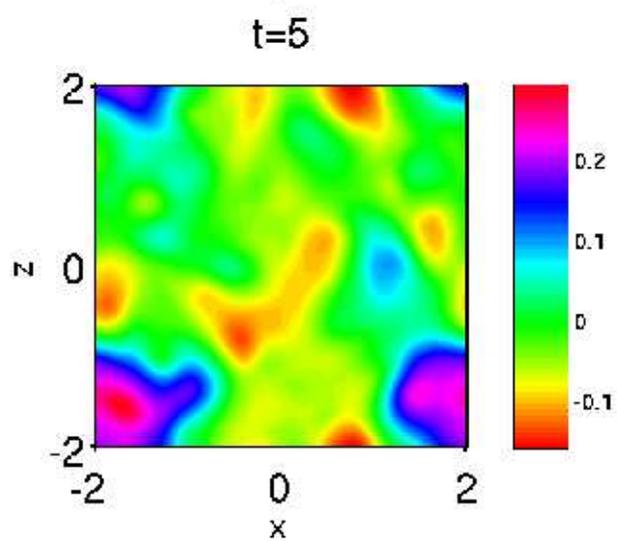
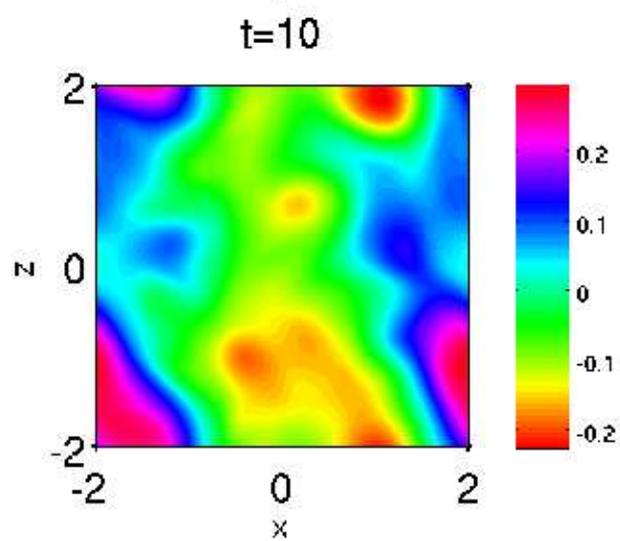
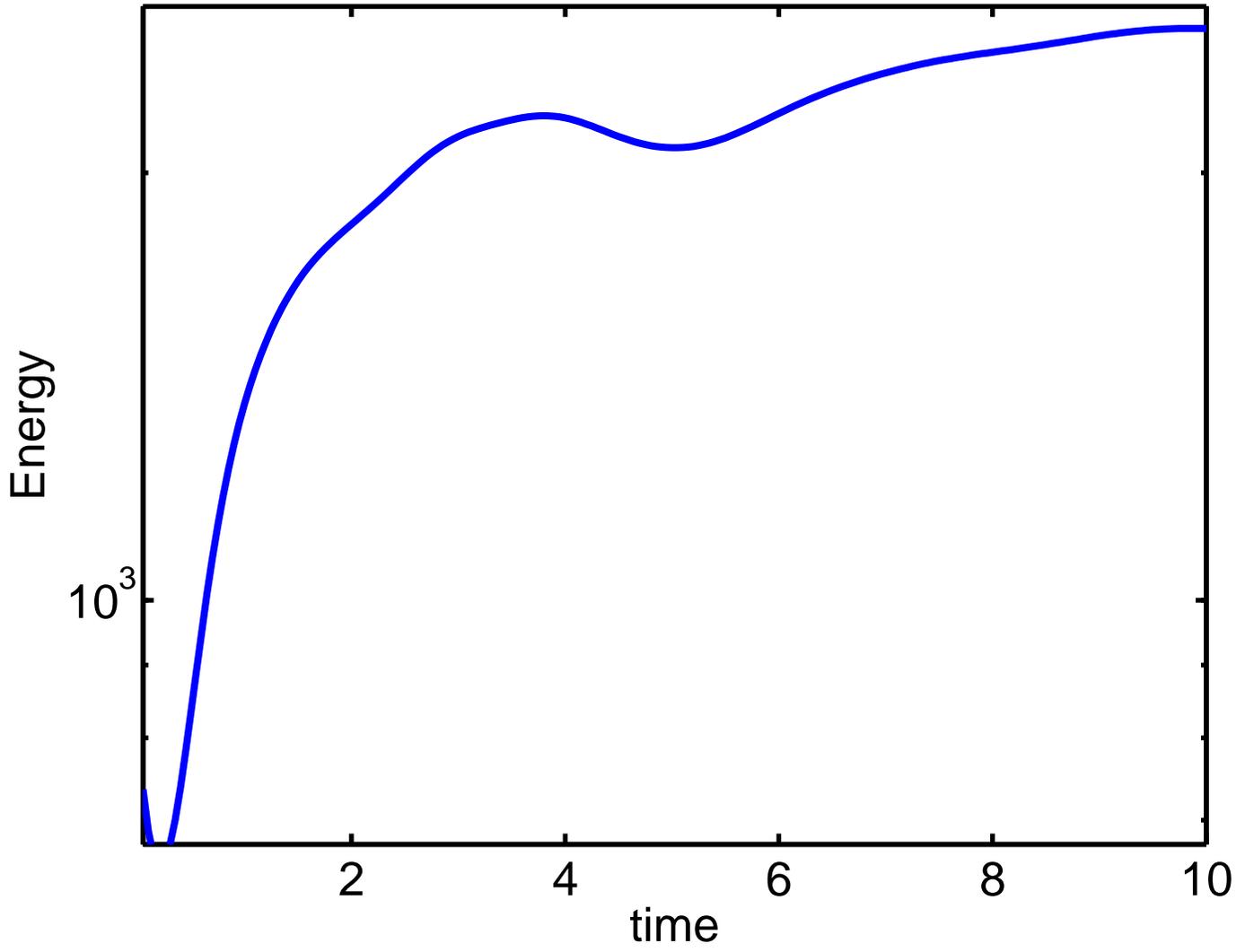
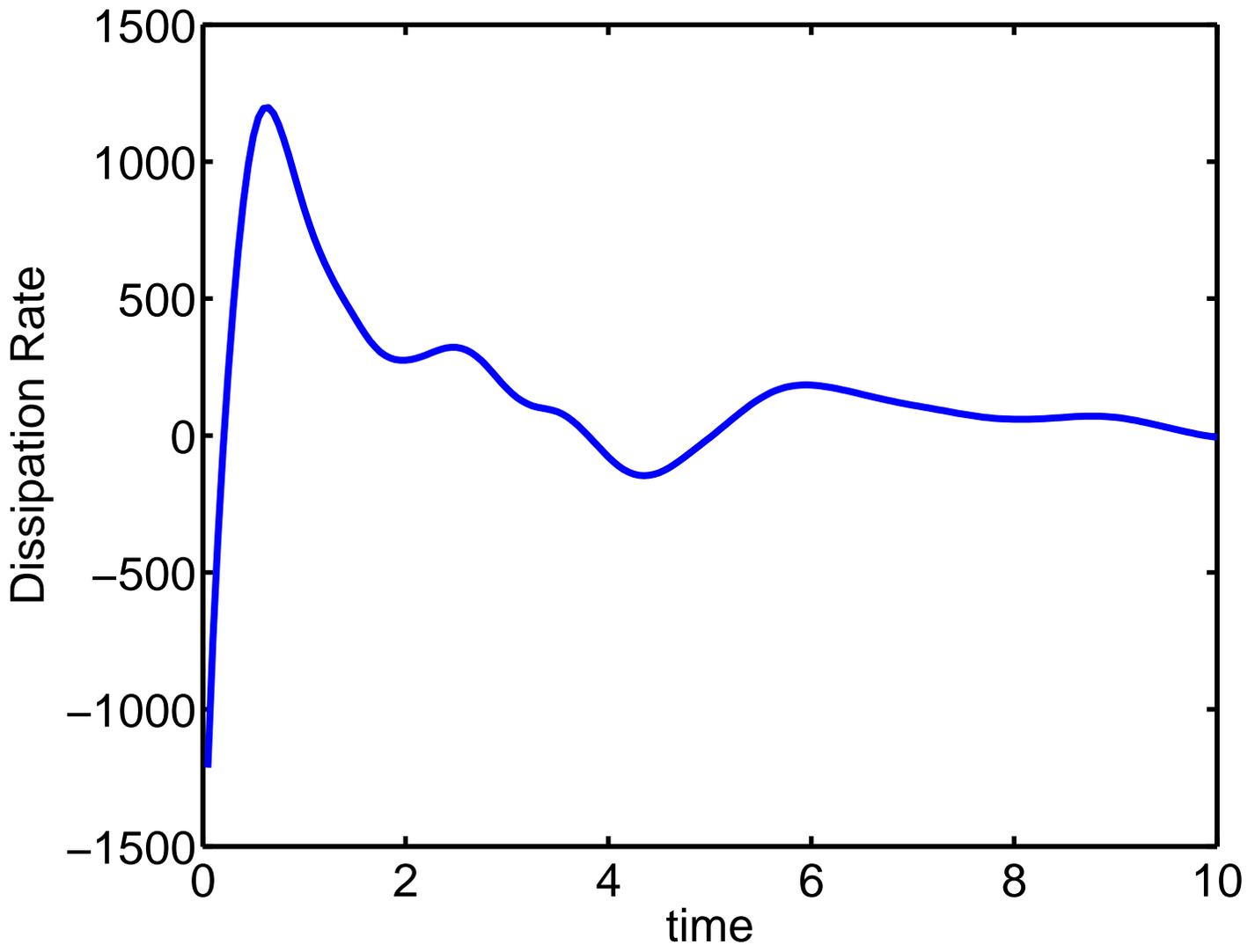
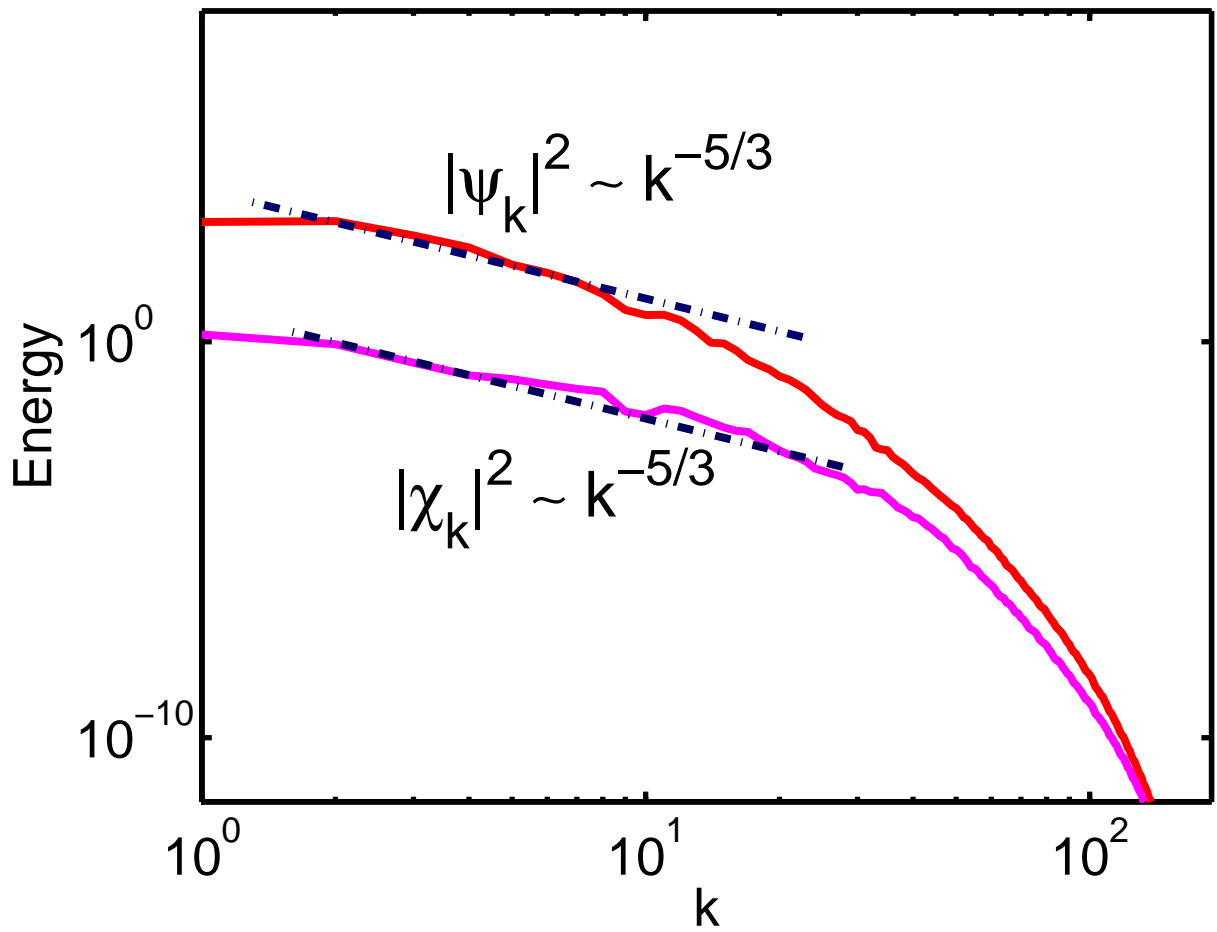
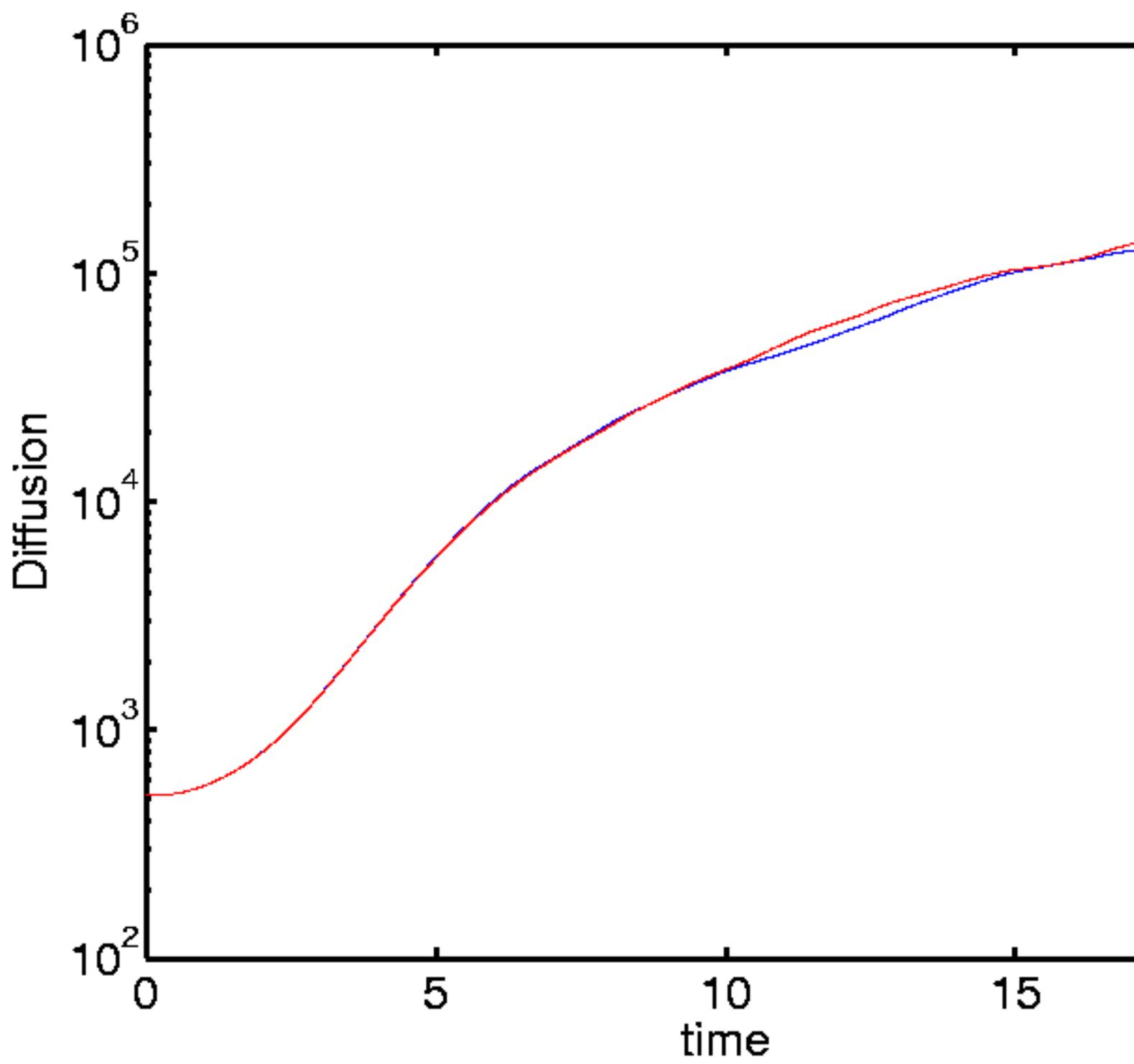
\end{document}